# Esports' Debut as a Medal Event at 2023 Asian Games: Exploring Public Perceptions with BERTopic and GPT-4 Topic Fine-Tuning

Tyreal Yizhou Qian
Louisiana State University
yqian@lsu.edu

Bo Yu
University of Minnesota
yub@umn.edu

Weizhe Li
Louisiana State University
wli49@lsu.edu

Chenglong Xu
Shanghai University of Sport
xuchenglong@sus.edu.cn

**Abstract**

*This study examined the public opinions of esports at the 2023 Asian Games and value co-creation during the event using an LLM-enhanced BERTopic modeling analysis. We identified five major themes representing public perceptions, as well as how major stakeholders co-created value within and beyond the esports ecosystem. Key findings highlighted the strategic use of social media marketing to influence public opinion and promote esports events and brands, emphasizing the importance of event logistics and infrastructure. Additionally, the study revealed the co-creation value contributed by stakeholders outside the traditional esports ecosystem, particularly in promoting national representation and performance. Our findings supported the ongoing efforts to legitimize esports as a sport, noting that mainstream recognition remains a challenge. The inclusion of esports as a medal event showcased broader acceptance and helped mitigate negative public perceptions. Moreover, contributions from non-traditional stakeholders underscored the value of cross-subcultural collaborations in esports.*

## 1. Introduction

Esports, once viewed as a niche interest, have achieved significant global recognition, drawing considerable attention and investments. Their inclusion as a medal event in the 2023 Hangzhou Asian Games (AG2023) exemplifies their evolving status and the broader acceptance of esports alongside traditional sports, signaling a groundbreaking moment in the sporting landscape (Goh, 2023). This integration highlights a transformative shift in the sports industry, mirroring wider societal trends towards digital entertainment. It challenges established notions of sports and prompts academic exploration into the underlying dynamics. Existing research has explored various aspects of esports, such as its competitive framework, economic sustainability, and societal impacts (Lu et al., 2024; Pizzo et al., 2022). However, much of this research focuses narrowly on a single esports genre and primarily reflects the perspectives of stakeholders deeply embedded within the esports ecosystem. This focus restricts broader inclusivity and may skew the understanding of esports, particularly concerning the interplay between esports and traditional sports, and the merging of these two realms.

The present study aimed to explore the diverse thematic discussions of esports at AG2023 on X. Unlike traditional methods that utilize surveys or experiments, this study analyzed user-generated content (UGC) from social media in various languages. Topic modeling, a prominent application of natural language processing (NLP), was employed to detect topics emerged from social media discourse. Specifically, we utilized BERTopic (Grootendorst, 2022) along with a large language model (LLM), GPT-4, to enhance topic representation (OpenAI, 2024). By employing advanced natural language processing (NLP) and machine learning (ML) techniques, our study attempted to enhance the understanding of public perceptions of esports and how value was co-created through esports at AG2023 from a stakeholder perspective, contributing to the ongoing discussion on the legitimization and sportification of esports (Heere, 2018; Scholz, 2020; Turtiainen et al., 2020). These findings can guide stakeholders in developing targeted strategies to enhance engagement, address public concerns, and provide insights on the sustainability of esports.

## 2. Background

### 2.1. Theoretical framework

Value co-creation, as outlined in the Service-Dominant (S-D) logic framework, provides a foundational lens for integrating esports into mainstream sports settings, such as the Asian Games 2023 (AG2023). S-D logic marks a paradigm shift from a goods-dominant to a service-dominant perspective, where value is co-created through interactions among multiple stakeholders who engage in resource exchange and mutual service provision (Kowalkowski, 2011; Vargo & Lusch, 2008, 2017). Scholars have explored the foundational premises of S-D logic within the marketing domain, focusing on areas such as branding, consumer culture, and marketing communication, while also examining the role of value propositions in these contexts (Vargo & Lusch, 2017).

The value-propositions concept, in particular, recognizes the central role of consumers in the value co-creation process, thereby emphasizing networked value propositions between consumers and multiple stakeholders (Cova & Salle, 2008; Frow & Payne, 2011; Kowalkowski, 2011; Vargo & Lusch, 2017). It is suggested that value co-creation occurs not only between producers and consumers but also through interactions with actors within both of their networks (Cova & Salle, 2008). In line with this, scholars have adopted the Stakeholder Theory (Freeman et al., 2010), which focuses on the relationships and mutual influences between organizations and stakeholders, to explore how value is collectively co-created within a network by aligning their interests and resources (Frow & Payne, 2011).

Previous studies have emphasized the interdependent relationship between organizations and their stakeholders, highlighting its significance in various contexts, including esports and traditional sports events (Jung et al., 2024; Kunz et al., 2022; Grohs et al., 2020; Parent, 2008). At AG2023, stakeholders including players, fans, sponsors, and event organizers played a pivotal role in the value creation process by actively participating in interactive experiences and sharing digital content, thereby enhancing the event's overall portrayal and success. Researchers have utilized S-D logic to investigate the interactions among esports stakeholders, confirming that their engagements at macro, meso, and micro levels are crucial for value co-creation within the esports ecosystem (Kunz et al., 2022; Roth et al., 2023). This collaborative effort among diverse service actors at esports events significantly shapes the co-creation of the live event experience (Jung et al., 2024).

The idea of value co-creation also highlights a significant shift in sports consumption, moving from a goods-dominant model to a service-oriented approach, in line with broader societal trends towards digital entertainment. The inclusion of esports as a medal event at AG2023 exemplifies this shift, demonstrating how digital and interactive entertainment forms are integrated into traditional sports frameworks. This integration underscores the evolving status of esports and illustrates the dynamic interaction between consumers and various stakeholders in co-creating the event's value. Understanding these dynamics is essential for comprehending how esports are not only being incorporated into global sporting events but also how they are reshaping traditional sports' stakeholder ecosystems. Therefore, our study leverages data from X to analyze discussions surrounding esports at AG2023, capturing the diverse perspectives of stakeholders involved. Based on this, we propose the following research questions (RQs):

RQ1: What themes emerge from the discussions about esports' debut at AG2023 on X?

RQ2: How do the discussions reflect value co-creation by relevant stakeholders?

### 2.2. Natural language processing and BERTopic

Existing research on esports has largely employed conventional data collection and analysis methods. These include qualitative panel discussions to understand organizational dynamics (Pizzo et al., 2022), quantitative surveys for hypothesis testing (Yu et al., 2022), and mixed methods approaches (Qian et al., 2020). While these methodologies are well-established, they come with certain inherent limitations. One major concern is the generalizability or external validity of the findings. For instance, qualitative interviews often involve small, non-representative samples. Likewise, measurement scales are typically predefined by researchers, which can limit respondents' ability to provide unexpected yet relevant insights (Hulland et al., 2018).

The proliferation of UGC provides a rich source of first-hand information, offering a way to overcome the limitations of traditional methodologies. To leverage the vast textual data from UGC, NLP, a subfield of artificial intelligence (AI) that trains computers to comprehend, interpret, and generate human language, has attracted significant attention (Qian et al., 2024; Rossi et al., 2024). Among its techniques, topic modeling, an unsupervised ML approach, has emerged as a powerful tool to discern key themes from large amounts of unstructured text. This method can uncover unanticipated and in-depth insights that smaller or more structured datasets might overlook (Mao et al., 2024). BERTopic, a recent innovation in topic modeling, utilizes transformer-based embeddings to improve the coherence and relevance of the identified topics (Grootendorst, 2022). Unlike traditional methods such as LDA topic modeling (Xu et al., 2024), BERTopic employs Bidirectional Encoder Representations from Transformers (BERT) to create dense, high-dimensional embeddings that more effectively capture contextual relationships between words (Egger & Yu, 2022).

More recently, the integration of LLMs like GPT-4 further amplifies BERTopic's ability to enhance topic representation by providing enriched semantic representations of text (Grootendorst, 2022). This integration enables more precise and nuanced topic extraction, as LLMs can comprehend and generate human-like text, capturing subtle nuances and context in language (Rossi et al., 2024). The combination of LLMs with BERTopic improves topic coherence by understanding the context of word usage, resulting in more meaningful and logically grouped topics. The relevance of the identified topics is also enhanced, as the advanced linguistic understanding provided by LLMs allows BERTopic to identify topics that are highly pertinent to the research context, even within complex and diverse datasets. Finally, the integration of BERTopic and LLMs enhances the method's ability to effectively process and analyze text in various languages, which is particularly advantageous for research involving international data sources. This multilingual functionality enables a more thorough examination of global discourse. In our study, the application of BERTopic combined with GPT-4 significantly improves our capability to capture the diverse thematic discussions on X surrounding esports at AG2023. This advanced method offers deeper insights into public perceptions and the complex nature of value co-creation by various stakeholders. Consequently, it contributes to a more comprehensive understanding of the ongoing discourse on the sportification of esports and its implications for the sports industry.

## 3. Methods

### 3.1. Data collection and pre-processing

As illustrated in Figure 1, we used Python to gather posts from X, employing libraries for web requests, parsing HTML, and handling URLs. We initiated an iterative keyword discovery method starting with "Asian Games" and "esports." Two researchers independently conducted open coding on 50 randomly chosen posts related to esports and the Asian Games on X, extracting frequently associated terms. This method, supported by collective discussions, was repeated until no new keywords emerged, signaling keyword saturation. Posts featuring any of the identified keywords were collected from August 1, 2022, to November 6, 2023. This timeframe ensures thorough coverage of AG2023, which took place from September 23 to October 8, 2023. Relevant metadata, such as timestamps, likes, and retweets, was also acquired during this process.

Next, the raw data was cleaned and preprocessed using the pandas and NLTK packages in Python. This preprocessing included eliminating duplicated posts, URLs, stopwords, punctuation, special characters, user mentions (e.g., "@username"), hyperlinks, and emojis. However, words were not subjected to stemming and lemmatization because contextual embeddings were used, which inherently understand the semantic relationships and variations in word forms (Grootendorst, 2022).

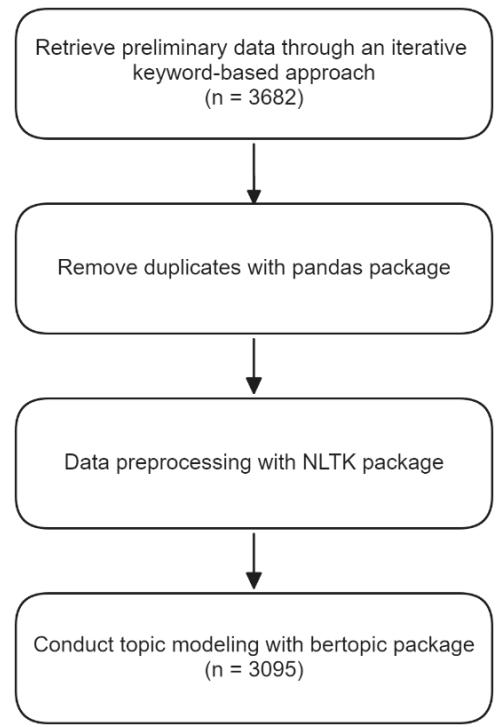

**Figure 1. Research flowchart.**

### 3.2. Topic modeling with BERTopic

BERTopic involves several key steps: embedding documents, reducing dimensionality, clustering embeddings, and representing topics. First, each document (tweet) was converted into an embedding representation—a vector of real numbers ($V \in R384$) that encapsulates the semantic meaning of the text. This step was executed using the pre-trained model "paraphrase-multilingual-MiniLM-L12-v2" (Reimers & Gurevych, 2019), which generates dense vector representations for multilingual paraphrases. Next, Uniform Manifold

Approximation and Projection (UMAP) was used to reduce the dimensionality of these embeddings, simplifying the high-dimensional data while preserving its structure. Default hyperparameters were applied during this process. After dimensionality reduction, Hierarchical Density-Based Spatial Clustering of Applications with Noise (HDBSCAN) was applied to cluster the similar embeddings using default settings. HDBSCAN identifies clusters of varying densities. The BERTopic model is particularly adept at handling the inherent noise in UGC due to its soft-clustering feature, which allows for overlapping clusters. The number of clusters is automatically determined by the algorithm, mitigating concerns over manually selecting the optimal number, as required in the Latent Dirichlet Allocation (LDA) approach. The final step involves generating topic representations for each cluster. Similar to the LDA approach, this is accomplished by qualitatively examining representative words and documents for each cluster derived from Class-based Term Frequency-Inverse Document Frequency (c-TF-IDF) (Grootendorst, 2022). However, unlike the LDA approach, BERTopic automatically determines the optimal number of topics. The results, including representative words and documents, can be reviewed and labeled either manually or using representation models. Some documents may be flagged as outliers; in this study, we employed topic-document distributions and c-TF-IDF representations to reassign topics to these outliers (Grootendorst, 2022).

To enhance topic representations, high-frequency yet uninformative stop words were removed from the cluster-specific documents. Additionally, we used GPT-4 to fine-tune our topic representations (Grootendorst, 2022). GPT-4, as one of the highest-performing LLMs, offers advanced NLP capabilities (OpenAI, 2024). To enable GPT-4 topic fine-tuning, we first generated a set of keywords and documents that best described each topic through our BERTopic model's c-TF-IDF calculation. These candidate keywords and documents were then input into GPT-4 (parameters: mode: complete, model: gpt-4-turbo-preview), prompting it to generate outputs that best fit the topic. As shown in Figure 2, we used [KEYWORDS] and [DOCUMENTS] tags in our prompts, indicating where to replace them with a topic's keywords and the top 10 most representative documents (nr_docs = 10), respectively. To ensure the credibility of the labels created, two independent researchers manually examined the labels to determine if they accurately reflected the keywords and documents. Our manual verification reached a consensus with a high agreement rate, demonstrating that the labels are both reliable and consistent with the content of the associated documents.

## 4. Results

To address RQ1, we conducted a BERTopic analysis to uncover the major themes of public discourses surrounding esports at AG2023. Topic representations based on c-TF-IDF and GPT-4 are shown in Table 1 and Table 2, respectively. Overall, our topic model resulted in 35 topics. Figure 3 displays the spatial distribution of topics and their interrelationships via a map-based interface.

To address RQ2, we employed a qualitative coding process similar to prior research (Liu et al., 2024) to group similar topics into macro-level themes. This allowed for meaningful interpretation through the lens of value co-creation and stakeholder theory. Given that our study was grounded in consumer interactions on X and aims to investigate the collective value consumers add to AG2023, we decided to focus on the value propositions perspective of S-D logic. This perspective highlights the pivotal role of consumers and emphasizes the network value propositions among consumers and various stakeholders (Cova & Salle, 2008; Frow & Payne, 2011; Kowalkowski, 2011; Vargo & Lusch, 2017).

Table 3 presents our analysis of 3,095 cleaned tweets discussing esports at AG2023, identifying five major themes about how value was co-created between esports consumers and other actors within their networks. The dominant theme, "Value Co-creation with Esports Community: Discussions on Competitions and Specific Titles," accounted for 26.59% of the tweets (n = 823), focusing on general esports events at AG2023 (Topics 3, 19, 24) and specific games like League of Legends (Topics 15, 18, 32), PlayerUnknown's Battlegrounds (Topic 16), and Arena of Valor (Topic 23). Tweets under this theme emphasized the collective value that consumers co-created through their interactions with esports communities, particularly those associated with different game titles, suggesting that various esports communities were significant stakeholders of AG2023 and actors of the consumer network. Closely following was "Value Co-creation with National Teams: National Representation and Performance," covering 26.53% of the discourse (n = 821), with emphasis on esports teams from China (Topics 10, 12, 26), South Korea (Topics 0, 20), India (Topics 7, 17, 21, 22), and the Philippines (Topic 4). This theme suggested that actors like national teams played crucial roles in the value co-creation process at AG2023, as national team performance and related media coverage significantly contributed to the perception of national pride and identity. "Value Co-creation with Athletes: Player Achievements", involving discussions on esports athletes as medalists and specific performances like that of South Korean LoL player Faker, including his military exemption, made up 21.20% (n = 656). Tweets in this theme indicated that esports athletes were also core actors of the AG2023 consumer network, as their individual stories and performances shaped the narrative of AG 2023. "Value Co-creation with Organizers: Event Logistics and Infrastructure" represented 19.87% (n = 615), detailing AG2023's schedule, venues, ancillaries, and partners (Topics 2, 8, 11, 14, 33, 28, 29, 25, 27). The organizers of AG2023 were one of the primary stakeholders and value co-creation actors, as the announcements of specific event logistics and ancillary events could enhance the perceived value of the event among consumers. The least prevalent theme was "Value Co-creation with Esports Entities: Recognition of Esports" at 5.81% (n = 180), which included debates on esports' status as a legitimate sport and medal event (Topics 5, 31, 34). Tweets in this theme contributed to the value co-creation between esports entities and consumers on the legitimacy of esports as a sport.

Table 1. c-TF-IDF representation.

| Topic | Count | c-TF-IDF |
|---|---|---|
| 0 | 173 | ['de', 'la', 'en', 'los', 'faker', 'korea', 'el', 'ruler', 'kanavi', 'keria'] |
| 1 | 128 | ['military', 'faker', 'gold', 'service', 'exemption', 'medal', 'win', 'bts', 'exempted', 'get'] |
| 2 | 113 | ['september', 'schedule', 'time', 'asian', 'games', 'worlds', 'lck', 'lpl', 'weeks', 'like'] |
| 3 | 301 | ['games', 'asian', 'play', 'lpl', 'lolesports', 'lplfanclub', 'watch', 'know', 'gt', 'lol'] |
| 4 | 103 | ['southeast', 'nd', 'cambodia', 'bang', 'malaysia', 'sibol', 'seagames', 'mobile', 'team', 'philippine'] |
| 5 | 104 | ['event', 'medal', 'esports', 'debut', 'hangzhou', 'medals', 'first', 'asiangames', 'make', 'sport'] |
| 6 | 174 | ['que', 'de', 'los', 'el', 'se', 'la', 'es', 'en', 'faker', 'lck'] |
| 7 | 96 | ['india', 'weareteamindia', 'teamindia', 'indiasports', 'indianesports', 'ptushaofficial', 'ianuragthakur', 'kirenrijiju', 'asiangames', 'media_sai'] |
| 8 | 90 | ['hangzhou', 'countdown', 'day', 'asiangames', 'esports', 'games', 'chinese', 'asian', 'centre', 'celebrate'] |
| 9 | 104 | ['gold', 'world', 'medal', 'sa', 'medalist', 'southeast', 'games', 'ang', 'asian', 'tunaynatabloidista'] |
| 10 | 100 | ['roster', 'team', 'china', 'jackeylove', 'coach', 'jiejie', 'knight', 'meiko', 'bin', 'camp'] |
| 11 | 83 | ['hearthstone', 'esports', 'removed', 'program', 'blizzard', 'games', 'chinese', 'asian', 'hangzhou', 'due'] |
| 12 | 77 | ['china', 'gold', 'first', 'ever', 'esports', 'hangzhou', 'valor', 'arena', 'malaysia', 'games'] |
| 13 | 143 | ['players', 'teams', 'game', 'best', 'asian', 'lpl', 'think', 'one', 'like', 'win'] |
| 14 | 50 | ['hangzhou', 'hangzhouasiangames', 'asiangames', 'nova', 'day', 'esports', 'chen', 'iqooind', 'aesf_official', 'tencentgames'] |
| 15 | 90 | ['league', 'legends', 'asian', 'th', 'game', 'games', 'riot', 'guy', 'nouveau', 'في'] |
| 16 | 53 | ['pubg', 'mobile', 'pubgmobile', 'roadtoasiangames', 'road', 'version', 'pubgmesports', 'asia', 'june', 'rdag'] |
| 17 | 65 | ['athletes', 'indian', 'sport', 'recognition', 'esports', 'ahead', 'industry', 'bats', 'year', 'next'] |
| 18 | 85 | ['lck', 'msi', 'worlds', 'lpl', 'year', 'finals', 'title', 'vs', 'final', 'spring'] |
| 19 | 96 | ['sports', 'events', 'competition', 'games', 'electronic', 'asian', 'cyber', 'hangzhou', 'online', 'world'] |
| 20 | 74 | ['team', 'ceremony', 'opening', 'national', 'hangzhou', 'korea', 'games', 'asian', 'esports', 'lol'] |
| 21 | 40 | ['training', 'kits', 'ioa', 'esfi', 'contingent', 'karmantikka', 'upcoming', 'conducts', 'kalyanchaubey', 'weareteamindia'] |
| 22 | 59 | ['fifa', 'fighter', 'street', 'dota', 'india', 'indian', 'online', 'asiangames', 'compete', 'esports'] |
| 23 | 53 | ['王者荣耀', 'original', 'asiangames', 'arena', 'mobilegame', 'aovnewskin', 'honorofkingslatam', 'aov', 'game', 'लक'] |
| 24 | 110 | ['di', 'esports', 'games', 'asian', 'ada', 'dan', 'cabor', 'yang', 'ini', 'ke'] |
| 25 | 54 | ['official', 'iqoo', 'gaming', 'phones', 'supplier', 'phone', 'vivo', 'th', 'esports', 'iqooind'] |
| 26 | 34 | ['national', 'dota', 'hangzhou', 'released', 'chinese', 'announced', 'coaches', 'team', 'esports', 'list'] |
| 27 | 27 | ['merchandise', 'hangzhou', 'official', 'store', 'look', 'batch', 'products', 'kings', 'honor', 'esports'] |

| Topic | Count | Words |
|---|---|---|
| 28 | 99 | ['game', 'peace', 'version', 'jacksonwang', 'jackson', 'wang', '王嘉尔', '잭슨', 'asian', 'teamwang'] |
| 29 | 49 | ['road', 'rdag', 'festival', 'aichi', 'nagoya', 'korizonesports', 'attend', 'macau', 'asia', 'asian'] |
| 30 | 107 | ['faker', 'south', 'like', 'chovy', 'asian', 'games', 'gaming', 'zeus', 'hyeok', 'get'] |
| 31 | 30 | ['martial', 'arts', 'indoor', 'aimag', 'proposed', 'medal', 'sport', 'official', 'في', 'riyadh'] |
| 32 | 35 | ['played', 'match', 'team', 'korea', 'hangzhou', 'legends', 'league', 'vietnam', 'ruler', 'pkl'] |
| 33 | 50 | ['hangzhou', 'venue', 'china', 'sports', 'center', 'events', 'centre', 'esports', 'meters', 'district'] |
| 34 | 46 | ['olympic', 'country', 'ioc', 'olympics', 'games', 'esport', 'asian', 'fires', 'esports', 'event'] |

**Table 2. GPT-4 representation.**

| Topic | Count | GPT4 |
|---|---|---|
| 0 | 173 | ["South Korea's Performance in League of Legends at Asian Games"] |
| 1 | 128 | ['Military exemptions for Korean esports athletes and celebrities.'] |
| 2 | 113 | ['Asian Games and Esports Competitions Schedule'] |
| 3 | 301 | ['Asian Games esports discussions'] |
| 4 | 103 | ['Philippine Achievements in Southeast Asian Games (Esports and Traditional Sports)'] |
| 5 | 104 | ['Esports as a Medal Event at the Asian Games'] |
| 6 | 174 | ['Asian Games and Professional Gaming Teams'] |
| 7 | 96 | ['Indian Esports Team at Asian Games'] |
| 8 | 90 | ['Countdown to Hangzhou Asian Games'] |
| 9 | 104 | ['Asian Games Gold Medal Winners'] |
| 10 | 100 | ['Team China Rosters for Asian Games'] |
| 11 | 83 | ['Removal of Hearthstone from Asian Games Esports Program'] |
| 12 | 77 | ['China wins first esports gold medal at Asian Games'] |
| 13 | 143 | ['Asian Esports Teams and Players'] |
| 14 | 50 | ['Hangzhou Asian Games Esports Events'] |
| 15 | 90 | ['League of Legends in Asian Games'] |
| 16 | 53 | ['PUBG Mobile Asian Games Events'] |
| 17 | 65 | ['Recognition of Indian Esports at Asian Games'] |
| 18 | 85 | ['International League of Legends Tournaments and Championships'] |
| 19 | 96 | ['E-sports Competitions at the Hangzhou Asian Games'] |
| 20 | 74 | ['Korean Esports Team at Hangzhou Asian Games'] |
| 21 | 40 | ['Training Kits for Indian Esports Athletes for Asian Games'] |
| 22 | 59 | ['Indian Esports at the Asian Games'] |
| 23 | 53 | ['Arena of Valor at Asian Games'] |
| 24 | 110 | ['Esports at the Asian Games'] |
| 25 | 54 | ['Official Esports Gaming Phone Supplier for the Asian Games'] |
| 26 | 34 | ['Chinese National Esports Teams for Hangzhou Asian Games'] |
| 27 | 27 | ['Hangzhou Asian Games official esports merchandise'] |
| 28 | 99 | ['Jackson Wang and Game Peace at Asian Games Esports'] |
| 29 | 49 | ['Road to Asian Games Esports Festival'] |
| 30 | 107 | ['South Korean gamer Faker at the Asian Games'] |
| 31 | 30 | ['Esports as an Official Medal Sport in Asian Indoor Games'] |
| 32 | 35 | ['League of Legends at Hangzhou Asian Games'] |
| 33 | 50 | ['Hangzhou Asian Games Esports Venue'] |
| 34 | 46 | ['Esports in the Asian and Olympic Games'] |

```
# System prompt describes information given to all conversations
system_prompt = """
<s>[INST] <<SYS>>
You are a helpful, respectful and honest assistant for labeling topics.
<</SYS>>
"""

# Example prompt demonstrating the output we are looking for
example_prompt = """
I have a topic that contains the following documents:

- Traditional diets in most cultures were primarily plant-based with a little meat on top, but with the rise of industrial style meat production and factory farming, meat has become a staple food.
- Meat, but especially beef, is the word food in terms of emissions.
- Eating meat doesn't make you a bad person, not eating meat doesn't make you a good one.
The topic is described by the following keywords: 'meat, beef, eat, eating, emissions, steak, food, health, processed, chicken'.
Based on the information about the topic above, please create a short label of this topic.
Make sure you to only return the label and nothing more.
[/INST] Environmental impacts of eating meat
"""

# Our main prompt with documents ([DOCUMENTS]) and keywords ([KEYWORDS]) tags
main_prompt = """
[INST]
I have a topic that contains the following documents:
[DOCUMENTS]

The topic is described by the following keywords: '[KEYWORDS]'.

Based on the information about the topic above, create a short label of this topic, ensuring comprehension across languages. Make sure you to only return the label and nothing more.
[/INST]
"""

prompt = system_prompt + example_prompt + main_prompt
```

**Figure 2. Prompt for GPT-4 topic fine-tuning.**

**Table 3. Maco-level themes of the topics.**

| Macro-level Themes | Topics | Count |
|---|---|---|
| Value Co-creation with National Teams: National Representation and Performance | 0, 4, 7, 10, 12, 17, 20, 21, 22, 26 | 821 |
| Value Co-creation with Organizers: Event Logistics and Infrastructure | 2, 8, 11, 14, 25, 27, 28, 29, 33 | 615 |
| Value Co-creation with Athletes: Player Achievements | 1, 6, 9, 13, 30 | 656 |
| Value Co-creation with Esports Entities: Recognition of Esports | 5, 31, 34 | 180 |
| Value Co-creation with Esports Community: Discussions on Competitions and Specific Titles | 3, 15, 16, 18, 19, 23, 24, 32 | 823 |

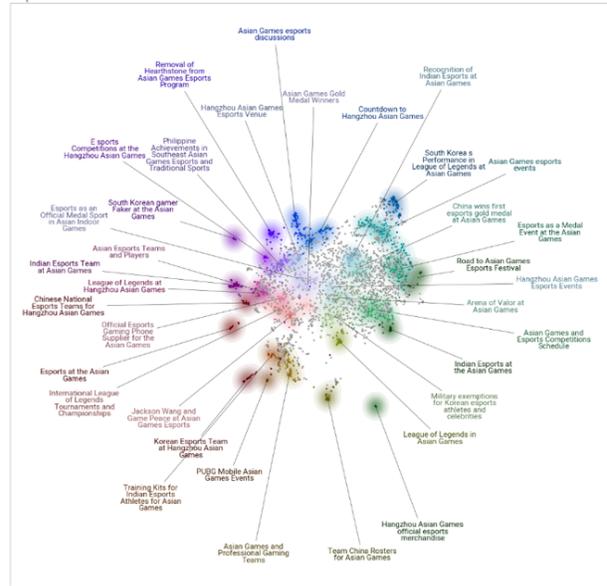

**Figure 3. Spatial distribution of topics.**

## 5. Discussion

The purpose of this study was to examine public perceptions of esports at AG2023 and explore how different stakeholders co-created value during the event. Through an LLM-enhanced BERTopic modeling analysis, we identified five major themes representing the public perceptions of AG2023 within and beyond the esports ecosystem. In alignment with the value propositions perspective of S-D logic, our primary findings suggested that the public discourse was centered around consumers and supported by various value propositions from multiple stakeholders (Cova & Salle, 2008; Frow & Payne, 2011; Kowalkowski, 2011; Vargo & Lusch, 2017). We identified several important stakeholders of AG2023, including national teams, esports entities, athletes, organizers, and esports communities, who played significant roles in the value co-creation with consumers.

Additionally, most identified themes supported the strategic use of social media marketing among stakeholders to influence public opinions and promote esports events and brands (Wang et al., 2024). Social media marketing of event logistics and infrastructure for esports at AG2023, such as the venue (Topic 33), schedules (Topic 2), ancillary events (Topic 8), and partners (Topic 28), was crucial for the organizing committee and its stakeholders as reflected by "Event Logistics and Infrastructure".

Another unique finding was the co-creation value contributed by stakeholders outside the esports ecosystem. Prior studies have noted the importance of interactions among internal and external stakeholders of the esports ecosystem (Scholz, 2020). Our study explored the advancement of esports at AG2023 as a result of the integration of resources and mutual benefits from both esports and traditional sports stakeholders. The theme of "National Representation and Performance" exemplifies this well. Esports stakeholders, such as national teams and athletes (Topic 7), promote their achievements, while traditional sports stakeholders, including the host country and participating nations (Topic 12), recognize team and individual success in esports to enhance national pride and demonstrate their soft power (Wong & Meng-Lewis, 2023).

The findings of our study supported the ongoing sportification attempts of esports. Previous research has recognized efforts to legitimize esports as a sport (Heere, 2018; Turtiainen et al., 2020). However, discussions under the theme "Recognition of Esports" indicated that achieving mainstream recognition continued to be an issue and required further progress. Similar to the inclusion of esports in the 2022 Commonwealth Games, recognition by traditional sports institutions, such as continental Games and national or regional sports bodies, could contribute to the legitimization of esports (Scholz, 2020). The inclusion of esports as a medal event demonstrated broader acceptance, contributing to the mitigation of negative public perceptions regarding esports (Topic 17). In addition, the broad acceptance of esports athletes being awarded medals at the Asian Games and the discussions about League of Legends player Faker's possible exemption from South Korean military service (Topic 30) highlighted the recognition of esports under the theme "Player Achievements."

One of the most intriguing findings was the contributions from individuals beyond the usual esports and sports stakeholders. A notable example includes the promotion of the theme song "Zone" for the game "Team for Peace," written and performed by Jackson Wang (Topic 28). Wang's involvement highlights AG2023's efforts to create intersections among various subcultures and demonstrates the participation of non-traditional stakeholders in the value co-creation for esports events. Promotion within subcultures has been suggested to transmit consumption values (Xu et al., 2024). However, the AG2023 case provides an extended perspective on the collaboration among subcultures, illustrating a broader approach to value co-creation in esports.

### 5.1. Theoretical implications

This study utilized the theoretical frameworks of value co-creation and stakeholder theory to analyze public discourse surrounding esports at AG2023 on X. Unlike previous research that focuses primarily on stakeholders within the esports ecosystem, we investigated the contributions of non-traditional esports stakeholders. Our findings offer two primary theoretical implications.

First, this study enhances the understanding of the value co-creation process both within and beyond the esports ecosystem. Consistent with research on traditional sports events, our findings lay the groundwork for identifying typical esports event stakeholders (e.g., teams, players) as well as non-traditional stakeholders (e.g., traditional sports governing bodies, pop culture celebrities). Furthermore, our study indicates that different stakeholders participate in the esports value co-creation process (Vargo & Lusch, 2008), each playing distinctive roles and contributing various values within the context of public discourse.

Second, our study contributes to the growing body of research examining public recognition of esports. Although AG2023 has adopted several esports as medal events, our findings suggest that public acceptance of esports still has a long way to go. These insights deepen our understanding of the sportification process of esports (Scholz, 2020; Turtiainen et al., 2020), particularly during the transitional phase of esports being recognized as an Olympic sport.

### 5.2. Practical implications

This study presents several empirical findings that can help esports practitioners develop effective strategies for leveraging events like Esports at AG2023. First, esports event organizers could enhance their social media marketing strategies by involving primary stakeholders, such as star players and teams, to attract more live event attendees. Engaging well-known players and successful teams in promotional activities can create a buzz and increase interest among fans, potentially leading to higher attendance and greater engagement at live events.

Second, it is essential for stakeholders to recognize the value co-creation by actors outside the esports ecosystem. This includes understanding the roles of various external stakeholders, such as sponsors, media, and broader gaming communities, in shaping the perception and success of esports events. When integrating esports into mega sports events, organizing committees should consider how these external stakeholders can contribute to the public recognition and legitimization of esports. This could involve collaborations with traditional sports organizations, leveraging mainstream media coverage, and engaging with broader community initiatives to enhance the visibility and acceptance of esports.

## 5.3. Limitations and future research

While our study offers notable theoretical contributions and empirical implications, we must acknowledge several limitations that could inform future research. First, our analysis is confined to the public discourse of esports at AG2023 on X. Given the restricted access to X in the host country, future researchers might explore primary social media platforms in the host country, such as Weibo and Douyin, to compare public discourse differences with our results.

Second, while BERTopic is a powerful tool for topic modeling, it does have some limitations (Egger & Yu, 2022). Smaller or noisy datasets can lead to less accurate topic models. BERTopic may also struggle with overlapping topics, where distinct topics share similar terms, potentially leading to less precise topic separation. It would be beneficial to incorporate other NLP methods, such as sentiment analysis and word embeddings, which could provide a more nuanced analysis of topics and overall trends. Future research should also consider integrating time-series analysis to enhance the depth and richness of the topic modeling results.

Finally, our findings are derived from interpretations of BERTopic results and are primarily descriptive. Future research could model these topics to empirically test the relationships between topics and engagement levels among stakeholders both within and outside the esports ecosystem, providing a deeper understanding of the value co-creation process.